\documentclass[aps,prl,twocolumn,groupedaddress,showpacs]{revtex4-1}
\usepackage{amsmath,amssymb}
\usepackage{graphicx}
\begin{document}

\title{Observation of a tricritical wedge filling transition in the 3D Ising model}
\author{A. Rodr\'{\i}guez-Rivas}
\affiliation{Departamento de F\'{\i}sica At\'omica, Molecular y Nuclear, \'Area de F\'{\i}sica Te\'orica,
Universidad de Sevilla, Apartado de Correos 1065, 41080 Sevilla (Spain)}
\author{J. M. Romero-Enrique}
\affiliation{Departamento de F\'{\i}sica At\'omica, Molecular y Nuclear, \'Area de F\'{\i}sica Te\'orica,
Universidad de Sevilla, Apartado de Correos 1065, 41080 Sevilla (Spain)}
\author{L. F. Rull}
\affiliation{Departamento de F\'{\i}sica At\'omica, Molecular y Nuclear, \'Area de F\'{\i}sica Te\'orica,
Universidad de Sevilla, Apartado de Correos 1065, 41080 Sevilla (Spain)}
\author{A. Milchev}
\affiliation{Institute for Physical Chemistry, Bulgarian Academy of Sciences, 1113 Sofia, Bulgaria}
\affiliation{Institut f\"ur Physik, Johannes Gutenberg-Universit\"at Mainz, Staudinger Weg 7, D-55099 Mainz, Germany}
\date{\today}

\begin{abstract}
In this Letter we present evidences of the occurrence of a tricritical filling transition for an Ising model in a linear
wedge.
We perform Monte Carlo simulations in a double wedge where antisymmetric fields act at the top and bottom wedges,
decorated with specific field acting only along the wegde axes. A finite-size scaling analysis of these simulations shows
a novel critical phenomenon, which is distinct from the critical filling. We adapt to tricritical filling 
the phenomenological theory which successfully was applied to the finite-size analysis of the critical filling in 
this geometry, observing good agreement between the simulations and the theoretical predictions for tricritical filling. 
\end{abstract}

%\keywords{}

\maketitle

Fluids adsorbed on micropatterned and sculpted solid substrates are
known to exhibit phase transitions which differ from those observed
at planar, homogeneous walls \cite{Gau,Rascon,Bruschi}. 
The simple 3D wedge geometry, characterized by a tilt angle $\alpha$, has been extensively studied in the past
\cite{Rejmer,Parry,Parry4,Parry2,Bednorz,Henderson,Henderson2,
Henderson3,Rascon2,JM1,JM2,JM3,Parry6,Nelson,Bruschi0,Bruschi,Bruschi2,Bruschi3,Milchev,Milchev2,Binder,JM4}.
Thermodynamic arguments \cite{Concus,Pomeau,Hauge} show that the wedge in presence of a saturated gas is
completely filled with liquid provided that the contact angle $\theta$
is less than the tilt angle $\alpha$. This transition may be first-order
or continuous (critical filling). For the latter, the characteristic lengthscales as the
averaged interfacial height $\ell_W$, the correlation length $\xi_y$
along the wedge axis and the interfacial roughness $\xi_\perp$ diverge continuously as
$\theta\to \alpha$ according to the power-laws
\begin{equation}
\ell_w\sim (\theta-\alpha)^{-\beta_W},\xi_y \sim
(\theta-\alpha)^{-\nu_y},\xi_\perp \sim (\theta-\alpha)^{-\zeta_W 
\nu_y}
\label{critexp}
\end{equation}
where the critical exponents for short-ranged forces take the values $\beta_W=1/4$, $\nu_y=3/4$ and $\zeta_W=1/3$. 
In a shallow wedge, corresponding to small $\alpha$, the conditions for observing continuous wedge filling transition are 
less restrictive than for critical wetting at planar walls, and critical filling may occur even if the walls of the wedge 
show first-order wetting \cite{Parry,Parry4}. In more acute wedges the situation is more complicated. For example recently 
it has been proposed that if the wedge is acute enough, the filling transition may become first-order even if
the wetting at the planar walls is continuous \cite{Parry6,Nelson}. 
Furthermore, a phenomenological theory which takes into account the breather mode interfacial fluctuations shows that 
the filling transition may be driven first-order by modifying the fluid-solid interactions close to the wedge bottom 
\cite{JM1,JM2,JM3,JM4}. The borderline between first-order and critical filling corresponds to a tricritical point, with
the strength of the fluid-solid interactions close to the wedge axis as a new relevant operator. However, along the
constant strength path the tricritical exponents Eq. (\ref{critexp}) are the same as in critical filling \cite{JM3},
so the observation of tricritical filling may be elusive.   

\begin{figure}[t]
\includegraphics[width=\columnwidth]{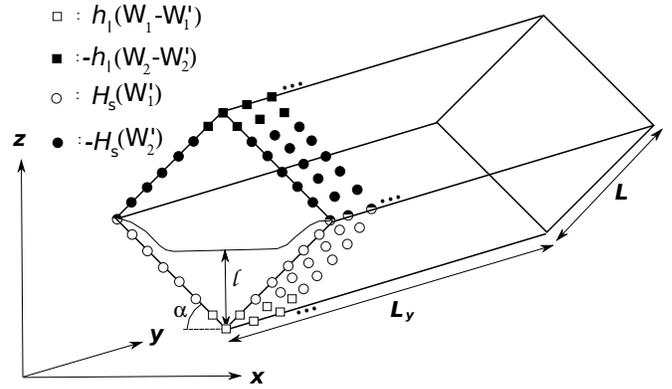}% Here is how to import EPS art
\caption{\label{figure1} Schematic picture of the modified antisymmetric double wedge Ising lattice. See text for
explanation.}
\end{figure}
Critical filling transitions have been observed unambiguously in computer simulation studies of the 3D Ising model 
\cite{Milchev,Milchev2,Binder,JM4}. The double wedge geometry with applied antisymmetric surface fields
is specially suitable to make a systematic 
finite-size characterization of the critical behavior of the filling transition, confirming 
the critical exponents (\ref{critexp}) obtained from interfacial Hamiltonian models \cite{Parry,Parry4}.
In this Letter we consider a suitable modification of this system in our search for the tricritical filling transition (see
Fig. \ref{figure1}). For this purpose, we perform Monte Carlo simulations for different box sizes, and the tricritical 
filling transition is located by the matching of the magnetization probability distribution functions (PDFs) 
with the theoretical prediction of the 
breather-mode model \cite{JM4}. This model considers that critical (and tricritical) filling phenomena at bulk coexistence 
can be understood using an effective pseudo-one-dimensional wedge Hamiltonian which accounts only for breather-mode 
excitations \cite{Parry,Parry4}: 
\begin{equation}
{\mathcal H}_W[\ell]=\int d y\left\{\frac{\Lambda(\ell)}{2}
\left(\frac{d\ell}{dy}\right)^2
+V_W(\ell)\right\}
\label{heff2}
\end{equation}
where $\ell(y)$ is the local height of the interface above the wedge bottom. For the double wedge geometry characterized
by a tilt angle $\alpha$, $0<\ell(y)<2L\sin\alpha$. The effective bending term $\Lambda(\ell)$ resisting fluctuations
along the wedge can be expressed as
\begin{equation}
\Lambda(\ell)= 
\frac{2\Sigma}{\tan \alpha}\min(\ell,2L\sin\alpha-\ell) 
\label{deflambda}
\end{equation}
For systems with short-ranged forces, the effective binding potential $V_W(\ell)$ has a short-ranged contribution,
which can be modeled as a contact potential of strength $-U$ at $\ell=0$ and $\ell=2L\sin\alpha$, and a 
long-ranged part which arises from the surface energy cost of forming a interfacial configuration: 
\begin{equation}
V_W (\ell)=\Lambda(\ell)\left(1-\frac{\cos\theta}{\cos\alpha}\right) 
\label{defvw}
\end{equation}
By setting $k_B T=1$ for convenience, the partition function $Z(\ell_b,\ell_a,L_y)$ can be represented as a 
path integral of $\exp(-{\mathcal H}_W)$ over all the paths which start for $y=0$ at an interfacial height $\ell_a$ 
and end at interfacial height $\ell_b$ for $y=L_y$ \cite{Burkhardt}.  
The presence of a position-dependent stiffness coefficient makes that some care is required in the definition of the 
partition function and its measure\cite{Bednorz}, similar to the factor-ordering problem in solid-state 
quantum mechanics when there is an effective position-dependent mass \cite{Thomsen,Chetouani}. We follow the prescription 
proposed in earlier work which provides a solution which is mathematical consistent and agrees with necessary thermodynamic 
requirements \cite{JM1,JM2,JM3}. The partition function $Z(\ell_b,\ell_a,L_y)$ can be expressed as \cite{Burkhardt}:
\begin{equation}
Z(\ell_b,\ell_a,L_y)=\sum_i \psi_i(\ell_b)\psi_i^*(\ell_a)
\textrm{e}^{-E_i L_y}
\label{partfunc4}
\end{equation}
where $\{\psi_i(\ell)\}$ is a complete orthonormal set of eigenfunctions with associated eigenvalues $E_i$
of the Hamiltonian operator $H_W$, 
\begin{equation}
H_W \equiv -\frac{1}{2}\frac{\partial}
{\partial \ell_b} \left[\frac{1}{\Lambda(\ell_b)}\frac{\partial}
{\partial \ell_b}\right]
+ V_W(\ell_b)+\tilde{V}_W(\ell_b)
\label{defhw}
\end{equation}
Here the term $\tilde{V}_W(\ell)$ is given by
\begin{equation}
\tilde{V}_W(\ell)=-\frac{1}{2\Lambda(\ell)}\left[\frac{3}{4}\left(\frac{\Lambda'(\ell)}
{\Lambda(\ell)}\right)^2-\frac{\Lambda''(\ell)}{2\Lambda(\ell)}\right]
\label{deftildev}
\end{equation}
where the prime denotes the derivative with respect to its argument. The short-ranged contribution to $V_W(\ell)$ leads
to boundary conditions for the eigenvalue problem. So, at $\ell=0$, we force the eigenfunctions to have the same 
short-distance expansion as in the infinite wedge situation \cite{JM3,JM4}
\begin{equation}
\psi (\ell) \sim \sqrt{\ell}\pm \frac{\Gamma[-1/3] 3^{-2/3}}{\Gamma[1/3]} \frac{\ell^{3/2}}{\xi_u} 
\label{sde}
\end{equation}
where $\xi_u\sim |U-U_{tc}|^{-1}$ is a characteristic
length associated with the contact potential at the wedge bottom and the positive and negative signs corresponds
to $U>U_{tc}$ (first-order filling) and $U<U_{tc}$ (critical filling), respectively, $U_{tc}$ being the tricritical
value \cite{JM3}. An analogous boundary condition is applied at the upper boundary $\ell=2L\sin\alpha$, substituting 
$\ell$ by $2L\sin\alpha-\ell$ in Eq. (\ref{sde}). 
%For critical filling, $\xi_u \to 0$ , so $\psi (\ell) \sim \ell^{3/2}$. 

For the case of periodic boundary conditions the interfacial partition function in the double wedge $Z_p$ can be obtained as
\begin{equation}
Z_{p}(L,L_y)=\int_{0}^{2L\sin \alpha} Z(\ell,\ell,L_y) d\ell=\sum_{i} e^{-E_{i}L_y}
\label{defz2}
\end{equation}
and the PDF for the interfacial height as \cite{JM4}
\begin{equation}
P_W(\ell,L,L_y)= \frac{\sum_{i} |\psi_i(\ell)|^2 e^{-E_{i}L_y}}{\sum_{i} e^{-E_{i}L_y}}
\label{defpw2}
\end{equation}
At the tricritical point (i.e. $\theta=\alpha$, $\xi_u\to \infty$), 
the eigenfunctions are alternating even and odd functions with respect to $\ell=L\sin\alpha$, which 
have in the interval $[0,L\sin\alpha]$ the expression 
\begin{equation}
\psi_i(\ell)\propto \sqrt{\ell}\left[\textrm{Ai}\left((-4\epsilon_i)^{1/3}\ell\right)
+\frac{1}{\sqrt{3}}\ \textrm{Bi}\left((-4\epsilon_i)^{1/3}\ell\right)\right]
\label{psi2}
\end{equation}  
where $\textrm{Ai}(x)$ and $\textrm{Bi}(x)$ are Airy functions, and the reduced eigenvalue 
$\epsilon_i=\Sigma E_i/\tan \alpha$.
Note that the eigenfunctions $\psi_i(\ell)$ satisfy the boundary conditions Eq. (\ref{sde}) for $\ell=0$ and the analogous 
boundary condition at $\ell=2L\sin\alpha$. On the other hand, the even eigenfunctions show a kink at $\ell=L\sin\alpha$, 
consequence of the Dirac-delta term $\Lambda''(x)$ in $\tilde{V}_W$ (see Eqs. (\ref{deflambda}) and (\ref{deftildev})), 
while the odd eigenfunctions vanish at $\ell=L\sin\alpha$. Thus the eigenvalues can be obtained 
by imposing the vanishing at $\ell=L\sin\alpha$ of the eigenfunction (odd case) or the 
derivative of $\psi(\ell)/\sqrt{\ell}$ (even case, see Eq. (21) in Ref. \cite{JM4}).
So, the associated eigenvalues have the expression $E_i=-\tan \alpha x_i^3/4\Sigma (L\sin\alpha)^3$, where $x_i$ are the 
solutions of the transcendental equations:
\begin{equation}
-\frac{1}{\sqrt{3}}=
\begin{cases}
\frac{\textrm{Ai}'(x)}{\textrm{Bi}'(x)} & \psi(\ell) \textrm{ even} \\
\frac{\textrm{Ai}(x)}{\textrm{Bi}(x)} & \psi(\ell) \textrm{ odd}
\end{cases} 
\label{bc2-3}
\end{equation}
which can be solved numerically or graphically. Table \ref{table1} shows the first few solutions to Eq. (\ref{bc2-3}). 
\begin{table}[t]
  \caption{\ First solutions of Eq. (\ref{bc2-3}).\label{table1}}
  \begin{ruledtabular}
  \begin{tabular}{ccc}
    $i$ & $x_i$ & Eigenfunction parity \\
    \hline
$0$ & $  0.0000$ & Even \\
$1$ & $-1.9864$ & Odd \\
$2$ & $-2.9488$ & Even \\
$3$ & $-3.8253$ & Odd \\
$4$ & $-4.5781$ & Even \\
$5$ & $-5.2956$ & Odd \\
$6$ & $-5.9503$ & Even \\
$7$ & $-6.5843$ & Odd \\
$8$ & $-7.1779$ & Even \\
  \end{tabular}
  \end{ruledtabular}
\end{table}

In order to compare these predictions with the existing Ising model computer simulation
results, we must convert the dependence on the interfacial height into an appropriate microscopic observable. 
For Ising model this is the  local magnetization density $m$ in the vertical plane at position $y$ along the wedge, 
which in the breather-mode picture can be related to $\ell$ as \cite{Albano,JM4}:
\begin{equation}
\frac{m}{m_b}=\begin{cases}
\left(\frac{\ell}{L\sin \alpha}\right)^2-1 & ;0<\ell<L\sin\alpha \\
1-\left(\frac{2L\sin\alpha-\ell}{L\sin \alpha}\right)^2 & ;L\sin\alpha < \ell <
2L\sin \alpha 
\end{cases}
\label{defm2d}
\end{equation}
where $m_b>0$ is the bulk magnetization density. The magnetization PDF is then related to the interfacial height
PDF via $P_W(m)=P_W(\ell(m))|d\ell/dm|$. At the tricritical filling transition, the magnetization PDF has an expression: 
\begin{widetext}
\begin{equation}
P_W(m)=
\frac{\sum_{i=0}^\infty {N_i^2
\left[\textrm{Ai}\left(-x_i \sqrt{1-\frac{|m|}{m_b}}\right)
+\frac{1}{\sqrt{3}}\textrm{Bi}\left(-x_i \sqrt{1-\frac{|m|}{m_b}}\right)\right]^2 
e^{\kappa x_i^3\frac{L_y}{L^3}}}}{\sum_{i=0}^\infty {e^{\kappa x_i^3
\frac{L_y}{L^3}}}}
\label{theoreticalpdf}
\end{equation}
\end{widetext}
where $N_i$ are the eigenfunction normalization factors and $\kappa=\tan \alpha/(4\Sigma \sin^3 \alpha)$.
Note that the PDF exactly at tricritical filling, as well as at critical filling \cite{JM4}, does not depend on $L$ and 
$L_y$ independently, but via the scaling combination $L_y/L^3$, i.e. $P_W(m,L,L_y)=P_W(m,L_y/L^3)$,
in agreement with previous scaling arguments \cite{Milchev,Milchev2,Binder}. For $L^3/L_y=0$, the magnetization PDF
over the interval $[-m_b,m_b]$ is flat, and as $L^3/L_y$ increases, the magnetization PDF becomes
bimodal, where the most-probable magnetization density in each section corresponds to $m=\pm m_b$.

The Monte Carlo simulations are performed for a nearest neighbor Ising model (isomorphic to a lattice gas)
on a simple cubic lattice with linear dimensions $L\times L \times L_y$ (in lattice spacing units):  
$19\times 19 \times 16$, $24\times 24 \times 37$, 
$34\times 34 \times 122$ and $44\times 44 \times 294$, where the ratio $L_y/(L-4)^3$ is approximately equal to 
$0.0046$. Periodic boundary conditions are applied along the $y$ direction, 
and in the remaining boundaries free boundary conditions are applied. On two of the surfaces which form the wedge
$W_1$, a linear field $h_l$ acts on the spins, and their nearest-neighbors, along the wedge tip,
and a surface fields $+H_s$ for the remaining spins in $W_1$. On the other wedge $W_2$ an opposite linear field $-h_l$ 
is applied for spins along or nearest-neighbor the wedge and a surface field $-H_s$ otherwise (see Fig. \ref{figure1}).  
We set the conditions which were previously considered for the critical filling characterization 
\cite{Milchev,Milchev2,Binder,JM4}, i.e. $\beta J \equiv J/k_B T =1/4$, and the surface exchange constant $J_s=J/2$. 
We take $J$ as the energy unit. The Hamiltonian of the Ising model is
\begin{widetext}
\begin{equation}
\beta H=-\frac{1}{4}\sum_{\langle i,j\rangle_{bulk}} S_i S_j 
-\frac{1}{8} \sum_{\langle i,j\rangle \in W_1 \cup W_2} S_i S_j
-\beta H_s \sum_{i\in W_1'} S_i + \beta H_s \sum_{i\in W_2'} S_i
-\beta h_l \sum_{i\in W_1-W_1'} S_i + \beta h_l \sum_{i\in W_2-W_2'} S_i
\label{ising}
\end{equation}
\end{widetext}
Under these conditions, the simulation box is a double wedge characterized by
a tilt angle $\alpha=\pi/4$, the value of $m_b\approx 0.75$, and $\Sigma=\beta\sigma a^2 
\approx 0.0981$, with $\sigma$ being the interfacial tension of the Ising model and $a$ the lattice spacing \cite{Milchev2}.
These Monte Carlo simulations were performed by using the standard Metropolis algorithm \cite{Metropolis}. 
The quantities we are interested in equilibrate quite slowly, so we considered runs of 
order of $10^8$ sweeps, where a sweep is $L^2\times L_y$ attempted updates of a spin chosen at random.
In each simulation we evaluated the magnetization PDF, where $m$
is defined as the average over the spins at each slice which are neither nearest nor next-nearest neighbors to $W_1$ 
or $W_2$ in order to minimize the effect of the enhanced order close to the surfaces.
The values of $H_s$ are taken to be the apparent critical filling values for each box size and $h_l=H_s$ \cite{JM4}
(in all cases $H_s\approx 0.72$). Note that the filling transition always occurs for $\theta=\alpha$ and the value of 
$H_s$ determines the value of the 
contact angle $\theta$, regardless the value of $h_l$. However, we have used single-histogram reweighting techniques 
\cite{Ferrenberg,Ferrenberg2} to tune the dependence on $H_s$ close to the filling transition of the magnetization PDF 
at a given value of $h_l$.   

\begin{figure}[t]
\includegraphics[width=\columnwidth]{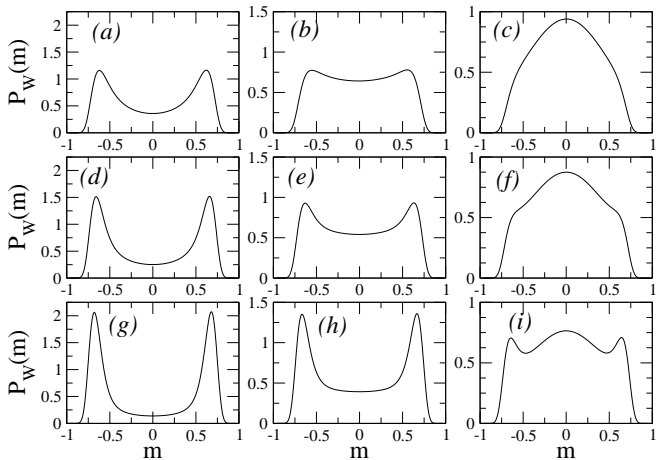}% Here is how to import EPS art
\caption{\label{figure2} Magnetization PDFs for $L=24$, $L_y=37$ and different values of $(H_s,h_l)$: 
(a) $(0.7084,0.7284)$, (b) $(0.7284,0.7284)$, (c) $(0.7484,0.7284)$, (d) $(0.7084,0)$, (e) $(0.7284,0)$, (f) $(0.7484,0)$,
(g) $(0.7084,-0.5)$, (h) $(0.7284,-0.5)$, and (i) $(0.7484,-0.5)$.}
\end{figure}

\begin{figure}[t]
\includegraphics[width=\columnwidth]{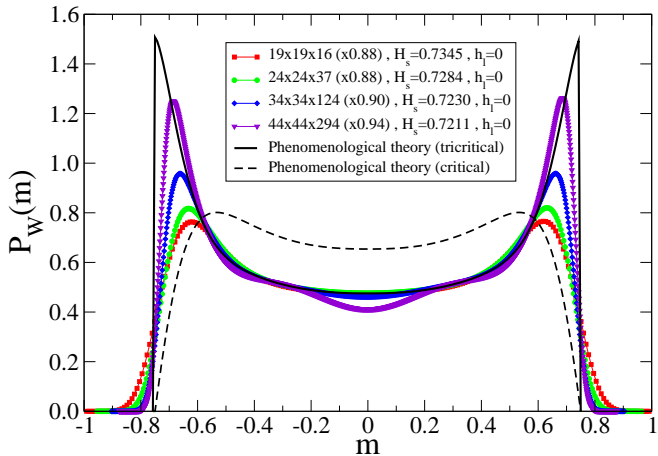}% Here is how to import EPS art
\caption{\label{figure3} (Color online) Plot of the magnetization PDFs for $h_l=0$ and: $H_s=0.7345$, 
$19\times 19\times 16$ (squares); $H_s=0.7284$, $24\times 24\times 37$ (circles); $H_s=0.7230$, $34\times 34\times 124$
(diamonds); and $H_s=0.7211$, $44\times 44\times 294$ (triangles). The continuous line corresponds to the breather-mode 
model predicted tricritical filling PDF, and the dashed line to the predicted critical filling PDF.}
\end{figure}

Figure \ref{figure2} shows typical magnetization PDFs for different values of $H_s$ and $h_l$. For all values of $h_l$,
the PDF is bimodal for values of $H_s$ well below the filling transition value, with maxima localized approximately at 
$\pm m_b$. On the other hand, if $H_s$ is well above the filling transition value, the PDF becomes unimodal with a single
maximum at $m=0$. Differences are observed when $H_s$ is around the filling transition value. 
For $h_l=H_s$, we reproduce the results already presented elsewhere \cite{JM4}, confirming that, under these
conditions, the filling transition is critical. Moreover, at the critical filling value of $H_s$ there is an excellent 
match with the predicted critical filling magnetization PDF from the phenomenological theory. 
If we set $h_l=-0.5$, a different scenario is observed. The magnetization PDFs for different
box sizes do not match the theoretical critical filling magnetization PDF. Actually, the location of the maxima is quite
insensitive to the value of $H_s$, and for large $H_s$ we observe a trimodal PDF with an additional maximum at $m=0$. 
By increasing $H_s$, the relative PDF height of the maxima at $m\approx \pm m_b$ with respect to the PDF value at $m=0$
decrease, until the former disappear. These observations are an indication that the filling transition may be first-order
for $h_l=-0.5$. We explored the values of $h_l$ between $h_l=-0.5$ and $h_l=0.72$ to locate the borderline between these
two scenarios, that we expect to be a tricritical point from our theoretical analysis. The procedure to locate the
tricritical point is as follows. As the simulation PDFs show tails for large $|m|$ (due to capillary fluctuations or other
irrelevant fluctuations), we match unnormalized PDFs (i.e. multiplied by an unknown factor to be determined in the
matching procedure) to the theoretical expression Eq. (\ref{theoreticalpdf}) in a magnetization window $|m|<m_{cut}$. 
For our simulations, we choose $m_{cut}=0.5$, finding the tricritical filling transition at $h_l\approx 0$. 
Figure \ref{figure3} shows the best matching magnetization PDFs for different simulation box sizes and $h_l=0$. The values
of $H_s$ correspond approximately to the transition values for critical filling, indicating that the filling transition
boundary is unaffected by the field $h_l$. On the other hand, the PDFs are clearly different from the critical filling PDF,
and as $L$ increases the two maxima converge to the theoretical tricritical PDF. This is the main result
of our paper, being a clear indication of the existence of a tricritical filling transition. Finally, it is worth 
noting that the magnetization PDF for the largest system seems to deviate from the theoretical prediction for small values
of $|m|$. This is also observed for critical filling \cite{unpublished}. We explain these discrepancies by the breakdown
of the breather-mode picture for small values of $|m|$ and large $L$. In fact, the analysis of typical snapshots shows
tilted configurations when $\ell\approx L/\sqrt{2}$, indicating that tilt and torsional modes \cite{Parry2} may be important under these conditions.  
  
In conclusion, we have found strong evidences that the filling transition can be driven tricritical by introducing a 
local field along the wedges which localizes the interface. To demonstrate this, we have performed Monte Carlo simulations
of the 3D Ising model in a double wedge geometry with applied antisymmetric surface fields and an additional field acting
along the wedges. A finite-size analysis of these simulations is in agreement with the predictions of the breather-mode
model for the tricritical filling. Although our study is restricted to the case of short-ranged forces,
it may be relevant for the case of dispersive forces, since the breather-mode model predicts that the filling transition
may also be driven first-order, with a critical end point as the borderline with the critical filling regime \cite{JM3}. 
We expect that the predictions for the latter case may be confirmed experimentally.

\begin{acknowledgments}
We thank Prof. K. Binder, Prof. A. O. Parry and Dr. C. Rasc\'on for discussions and reading of the manuscript. A. R.-R. 
thanks the Institut f\"ur Physik (Johannes Gutenberg-Universit\"at Mainz) 
for hospitality during a research stay where this work was begun. A. R.-R., J. M. R.-E. 
and L. F. R. acknowledge financial support from the Spanish Ministerio de Econom\'{\i}a y Competitividad through 
grants no. FIS2009-09326 and FIS2012-32455, and Junta de Andaluc\'{\i}a through grant no. P09-FQM-4938, all co-funded 
by the EU FEDER, and the Portuguese Foundation for Science and Technology under Contract No. EXCL/FIS-NAN/0083/2012.
\end{acknowledgments}

\bibliography{paper}

\end{document}